\documentclass[a4paper,11pt]{article}
\usepackage[usenames]{color}
\usepackage[a4paper,top=3cm,left=3cm,right=3cm,bottom=3cm]{geometry}
\usepackage{amsfonts,amssymb}
\usepackage{theorem}
\usepackage[dvips]{graphicx}

\def\Astarbl{\widetilde{\widehat A^*_\mu}}

\def\G{\Gamma}

\def\wG{\widehat \Gamma}

\newcommand{\evg}[1]{\color{green}{#1~}\color{black}}%

\def\s{\mathfrak{s}}

\begin{document}
\par\noindent
\rightline{FR-PHENO-2010-023}
\rightline{IFUM-932-FT}
\rightline{July 2010}

\vskip 2 truecm
\bf
\Large
\centerline{The Algebra of Physical Observables}
\centerline{in Nonlinearly Realized Gauge Theories}
\rm

\vskip 1.5 truecm
\large
\centerline{
Andrea Quadri \footnote{E-mail address: 
{\tt andrea.quadri@mi.infn.it}}}

\vskip 0.3 truecm
\normalsize
\centerline{Physikalisches Institut, 
Albert-Ludwigs-Universit\"at Freiburg}
\centerline{Hermann-Herder-Strasse 3, D-79104 Freiburg, Germany}
\centerline{{\em and}}
\centerline{Phys. Dept. University of Milan, 
via Celoria 16, I-20133 Milan, Italy }

\normalsize
\rm

\vskip 1.5  truecm
\centerline{\bf Abstract}
\vskip 0.3 truecm

\begin{quotation}
\noindent
We classify the physical observables in 
spontaneously broken 
nonlinearly realized
gauge theories in the recently proposed loopwise
expansion governed by the
Weak Power-Counting (WPC)  
and the Local Functional Equation. The latter controls
the non-trivial quantum deformation of the classical
nonlinearly realized gauge symmetry, to all orders
in the loop expansion.
The Batalin-Vilkovisky (BV) formalism is used.
We show that the dependence of the vertex functional
on the Goldstone fields is obtained via a canonical
transformation w.r.t. the BV bracket associated
with the BRST symmetry of the model.
We also compare the WPC with strict power-counting
renormalizability in linearly realized gauge theories.
In the case of the electroweak group
we find that the tree-level Weinberg relation 
still holds if power-counting renormalizability
is weakened to the WPC condition.
\end{quotation}
\newpage

\section{Introduction}

The theoretical understanding of the mass
generation mechanism in non-Abelian gauge theories,
which will be experimentally probed in the coming years
at the LHC, is a challenging open issue in today's
high-energy physics.

The spontaneous symmetry breaking realization
based on the Higgs mechanism \cite{higgs}
is a
sound option which leads to the phenomenologically successful
Standard Model (SM) of particle physics \cite{SM}, a theory
which is both physically unitary and power-counting
renormalizable.
In models based on the Higgs mechanism,
at least one additional physical scalar
particle is present in the perturbative spectrum.

Due to the lack of experimental evidence for the
Higgs resonance, other possibilities have
nevertheless been investigated.
Higher dimensional models have been
intensively studied 
\cite{hd}.
Higgsless models based on 
modified energy-dependent running coupling constants have been considered \cite{Moffat:2010jx}.

Chiral models 
have been proposed since a long time \cite{Appelquist:1980vg}.
They are formulated  in the presence of a (classical)
non-linearly realized non-Abelian gauge symmetry.
The perturbative treatment of these theories
is usually performed in the momentum expansion,
leading to the low-energy Higgsless effective field theory
of the chiral electroweak lagrangian \cite{EFT}.

More recently an approach based on the perturbative loop
expansion of models endowed with a nonlinearly realized 
gauge symmetry has been
investigated.
The discovery of the Local Functional Equation (LFE)
\cite{Ferrari:2005ii}
has provided a key tool 
in the program of taming the divergences
of the nonlinearly realized theories 
recursively in the loop number.
The LFE encodes
the invariance of the path-integral  Haar measure
under local nonlinearly realized gauge transformations.
It provides a consistent way to handle the
non-trivial deformation of the classical chiral symmetry
induced by radiative corrections \cite{Bettinelli:2007kc}.
The LFE enforces a hierarchy among 1-PI Green functions:
those containing at least one Goldstone field (descendant
amplitudes) are fixed in terms of amplitudes which do not
involve any Goldstone leg (ancestor amplitudes)
\cite{Ferrari:2005ii}.
%
Applications to the nonlinear sigma model in $D=4$
have been given in \cite{Ferrari:2005ii}-\cite{Ferrari:2005va}.
The massive nonlinearly realized SU(2) Yang-Mills theory has been
studied in \cite{Bettinelli:2007cy}-\cite{Bettinelli:2007eu}
and the nonlinearly realized electroweak (EW) model
has been formulated in \cite{Bettinelli:2008ey}-\cite{Bettinelli:2009wu}.

The LFE technique has also
found applications to nonlinearly realized
field transformations, like e.g. polar coordinates
in the complex free field theory in $D=4$ \cite{Ferrari:2009uj}.

The present paper is devoted to the cohomological
characterization of the algebra of physical observables
in the nonlinearly realized (non-anomalous) gauge theories,
in the framework of the quantization procedure based
on the LFE and the loop expansion.

In view of the availability of an all-orders mathematically
consistent formulation of (quantum deformed) gauge symmetry via
the LFE, it is very important to be able to classify all the physical observables in nonlinearly realized gauge models 
on symmetry grounds and to connect them
with their classical counterpart.

The appropriate framework for such a task is provided by the
Batalin-Vilkovisky (BV) formalism \cite{Batalin:1981jr},\cite{Gomis:1994he}.
For the sake of simplicity we work with the
nonlinearly realized SU(2) Yang-Mills theory, but the
results can be easily extended to more general
gauge groups.
We make use of cohomological
techniques \cite{homol} in order to analyze
the physical observables of the theory.
The dependence of the 1-PI vertex
functional on the Goldstone fields, dictated by the LFE,
turns out to be generated via a canonical transformation w.r.t. the
BV 
bracket 
induced
by the BRST symmetry.

The zero ghost number
 cohomology $H_0({\cal S}_{\wG})$ of the full linearized BV bracket ${\cal S}_{\wG}$
(which takes into account the effects of the deformation
of the nonlinearly realized gauge symmetry)
is shown to be isomorphic to the zero ghost number cohomology
$H_0({\cal S}_0)$
of the classical linearized BV bracket ${\cal S}_0$.
The latter is given 
for the massive SU(2) Yang-Mills theory
based on the nonlinearly realized gauge group
by the set of all possible 
global $SU(2)_R$-invariant polynomials
in the bleached variable $a_\mu$ (the  classical gauge-invariant combination
of the gauge field $A_\mu$ and of the Goldstone fields $\phi_a$
which reduces at $\phi_a=0$ to $A_\mu$) and its ordinary
derivatives which do not vanish on the tree-level equation of motion
for $a_\mu$.

However, not all these (integrated) operators are allowed in the tree-level
vertex functional. In fact a Weak Power-Counting (WPC)
condition holds 
\cite{Ferrari:2005va,Bettinelli:2007tq,Bettinelli:2008qn}
for the model at hand. The WPC
states that 
only a finite number of divergent ancestor
amplitudes exists order by order in the loop expansion.
The validity of the WPC  
provides in turn a very restrictive
selection criterion for the operators which can be introduced
in the tree-level vertex functional. 
Only the standard Yang-Mills field strength squared
plus the St\"uckelberg mass term 
are compatible with the WPC 
and all the symmetries of the theory 
\cite{Bettinelli:2007tq}.

The situation is more involved for the nonlinearly
realized SU(2) $\otimes$ U(1) EW model. There the
WPC predicts \cite{Bettinelli:2008ey,Bettinelli:2008qn} the same couplings in the 
gauge and fermionic matter sectors (at zero
Goldstone fields) as in the Standard
Model in the unitary gauge. It however allows for two independent mass
invariants in the vector meson sector, i.e. the tree-level
Weinberg relation $M_Z=M_W/c_W$ (where
$M_Z$ and $M_W$ are the $Z$ the $W$ masses
and $c_W$ is the cosine of the Weinberg angle) does not hold.

It should be emphasized  that in the subtraction
scheme controlled by the WPC and the LFE 
the existence of a second
mass invariant is intimately related to the nature
(linear or nonlinear) of the gauge group realization.
We will indeed prove that the tree-level
Weinberg relation still holds
in the presence of the linearly realized EW gauge symmetry if power-counting
renormalizability is weakened to the WPC condition.


\medskip
The paper is organized as follows.
In Sect.~\ref{sec:2} we discuss the 
nonlinearly realized SU(2) massive Yang-Mills theory and its
symmetries. The BV bracket is defined in Sect.~\ref{sec:3}.
The master equation is derived in the same Section.
In Sect.~\ref{sec:4} we study the cohomologies
in ghost number zero of the quantum and classical
linearized BV brackets and prove that they are isomorphic.
In Sect.~\ref{sec:5} we compute the
cohomology in ghost number zero for the
classical linearized BV bracket.
In Sect.~\ref{sec:6} we show that the bleached
variables (invariant under the linearized LFE) 
are generated via a canonical transformation.
In Sect.~\ref{sec:7} we compare the allowed interaction terms,
compatible with the WPC, in the framework of the linearly 
vs. the nonlinearly realized SU(2) gauge theory.
Sect.~\ref{sec:8} extends this analysis to the
SU(2) $\otimes$ U(1) gauge group.
Conclusions are finally given in Sect.~\ref{sec:9}.

\section{The Model and its Symmetries} \label{sec:2}

We consider pure massive Yang-Mills theory based on the nonlinearly
realized SU(2) gauge group \cite{Bettinelli:2007tq}.
By imposing the relevant symmetries of the theory
(Slavnov-Taylor (ST) identity, LFE,
ghost equation, global SU(2)$_R$ invariance)
and the requirement of
the WPC a unique tree-level
vertex functional arises \cite{Bettinelli:2007tq}
\begin{eqnarray}
\G^{(0)} & = & \Lambda^{D-4} \int d^Dx \, \Big (
-\frac{1}{4g^2} G_{a\mu\nu}G^{\mu\nu}_a + \frac{M^2}{2}
(A_{a\mu} - F_{a\mu})^2 \nonumber \\
& & ~~ + B_a D^\mu[V](A-V)_{a\mu} -
            \bar c_a (D^\mu[V] D_\mu[A] c)_a
          + \bar c_a (D^\mu[A] \Theta_\mu)_a 
       \nonumber \\
& & ~~ + A^*_{a\mu} \mathfrak{s} A_a^\mu 
       + \phi_0^* \mathfrak{s} \phi_0 
       + \phi_a^* \mathfrak{s} \phi_a  - c_a^* \mathfrak{s} c_a + K_0 \phi_0 \Big ) \, .
\label{os.eq.1}
\end{eqnarray}
The gauge bosons acquire a mass via the St\"uckelberg 
mechanism \cite{stueck_orig,stuck}.

In the above equation
$G_{a\mu\nu}$ is the non-Abelian field strength
\begin{eqnarray}
G_{a\mu\nu} = \partial_\mu A_{a\nu} - \partial_\nu A_{a\mu} + 
\epsilon_{abc} A_{b\mu} A_{c\nu} \, .
\label{os.eq.5}
\end{eqnarray}
The covariant derivative is defined as 
\begin{eqnarray}
D_\mu[A]_{ab} = \delta_{ab} \partial_\mu + \epsilon_{acb} A_{c\mu} \, .
\label{cov.dev.}
\end{eqnarray}
$g$ denotes the gauge coupling constant.

The BRST differential $\mathfrak{s}$ acts as follows on the
fields of the theory
\begin{eqnarray}
&& \s A_{a\mu} = \partial_\mu c_a + \epsilon_{abc}
A_{b\mu} c_c \, , ~~~~ \s c_a = -\frac{1}{2} \epsilon_{abc} c_b c_c \, ,
\nonumber \\
&& \s \phi_a = \frac{1}{2} \phi_0 c_a + \frac{1}{2}
\epsilon_{abc} \phi_b c_c \,  , ~~~ \s\phi_0 = - \frac{1}{2} \phi_a c_a \, , 
\nonumber \\
&& \s \bar c_a = B_a \, , ~~~~ \s B_a = 0 \, .
\label{os.eq.6}
\end{eqnarray}
$c_a$ are the ghost fields, $\bar c_a$ the antighost fields and
$B_a$ the Nakanishi-Lautrup \cite{naklautrup,Piguet:1995er} multiplier fields. $\phi_0$ is the solution of the nonlinear constraint
\begin{eqnarray}
\phi_0^2 + \phi_a^2 = v^2 \, , ~~~~~ \phi_0 =\sqrt{v^2 - \phi_a^2} \, .
\label{nonlinconst}
\end{eqnarray}
For the sake of simplicity and conciseness we have adopted
the Landau gauge.
The extension to an arbitary 't Hooft gauge is discussed in \cite{Bettinelli:2007eu}.

$V_{a\mu}$ is the background connection necessary for the implementation
of the LFE.
It is paired in the usual fashion \cite{bgf} with the
background ghost $\Theta_{a\mu}$ into a BRST doublet
\cite{Quadri:2002nh}
\begin{eqnarray}
\s V_{a\mu} = \Theta_{a\mu} \, , ~~~~ \s \Theta_{a\mu} = 0 \, .
\label{os.eq.7}
\end{eqnarray}
$K_0$ is the scalar source coupled to the nonlinear constraint
$\phi_0$ in eq.(\ref{os.eq.1}).
While the background ghost $\Theta_{a\mu}$
is not needed in order to formulate the LFE and the ST
identity, it is an expedient technical tool in order to show
that physical observables do not depend on the
background connection $V_{a\mu}$, as explained
in Refs.~\cite{bgf}.

The antifields $A_{a\mu}^*,\phi_a^*,c_a^*$ and $\phi_0^*$
are external sources coupled to the nonlinear BRST variations 
\cite{Gomis:1994he,zj}
of the 
corresponding fields $A_{a\mu}, \phi_a, c_a$ and 
$\phi_0$.
The antifield $c_a^*$ has an extra minus sign
w.r.t. the conventions adopted in \cite{Bettinelli:2007tq}.
This choice turns out to be convenient in the definition of the
Batalin-Vilkovisky bracket in Sect.~\ref{sec:3}.

Notice that the presence of the scalar source $K_0$
(required for the formulation of the LFE) forces the
introduction of the antifield $\phi_0^*$ in order to derive
the ST identity, despite the fact that $\phi_0$ is not an elementary
field. 
The source $K_0$ is paired with $\phi_0^*$ into a BRST doublet as follows
\begin{eqnarray}
\s \phi_0^* = - K_0 \, , ~~~~ \s K_0 = 0 \, .
\label{os.eq.8}
\end{eqnarray}
$\Lambda$ is a mass scale for continuation in $D$ dimensions.
In the present paper we choose to factor out $\Lambda$ in front of the
full tree-level vertex functional $\G^{(0)}$. This will
simplify the notations in the discussion of
the BV formalism in Sect.~\ref{sec:3}.

The following functional identities hold for the 1-PI vertex functional
$\G$~\cite{Bettinelli:2007tq}:
\begin{itemize}
\item the ST identity
\begin{eqnarray} 
{\cal S}(\G) & = & \int d^Dx \, \Bigg [ \frac{1}{\Lambda^{(D-4)}} \Big ( 
                   \frac{\delta \G}{\delta A_{a\mu}^*}
                   \frac{\delta \G}{\delta A_a^\mu}  
                   +
                   \frac{\delta \G}{\delta \phi_a^*}
                   \frac{\delta \G}{\delta \phi_a} 
                   -
                   \frac{\delta \G}{\delta c_a^*}
                   \frac{\delta \G}{\delta c_a} \Big ) \nonumber \\
             &   & ~~~~~~~~ + B_a \frac{\delta \G}{\delta \bar c_a} 
                   + \Theta_{a\mu} \frac{\delta \G}{\delta V_{a\mu}} 
                   - K_0 \frac{\delta \G}{\delta \phi_0^*} \Bigg ] = 0 
\label{st.1}
\end{eqnarray}
\item the LFE
\begin{eqnarray}
\!\!\!\!\!\!\!\!\!\!
{\cal W}_a (\G) & = & -\partial_\mu \frac{\delta \G}{\delta V_{a\mu}}
                             + \epsilon_{abc} V_{c\mu} \frac{\delta \G}{\delta V_{b\mu}} 
                             -\partial_\mu \frac{\delta \G}{\delta A_{a\mu}}
                             + \epsilon_{abc} A_{c\mu} \frac{\delta \G}{\delta A_{b\mu}} 
                      \nonumber \\
                &   & +\frac{1}{2 \Lambda^{D-4}} \frac{\delta \G}{\delta K_0}\frac{\delta \G}{\delta \phi_a}
                      +\frac{1}{2} \epsilon_{abc} \phi_c \frac{\delta \G}{\delta \phi_b} \nonumber \\
                &   & + \epsilon_{abc} B_c \frac{\delta \G}{\delta B_b} 
                      + \epsilon_{abc} \bar c_c \frac{\delta \G}{\delta \bar c_b} 
                      + \epsilon_{abc} c_c \frac{\delta \G}{\delta c_b} \nonumber \\
                &   & + \epsilon_{abc} \Theta_{c\mu} \frac{\delta \G}{\delta \Theta_{b\mu}}
                      + \epsilon_{abc} A^*_{c\mu} \frac{\delta \G}{\delta A^*_{b\mu}}
                      + \epsilon_{abc} c^*_c \frac{\delta \G}{\delta c^*_b} \nonumber \\
                &   & + \frac{1}{2} \phi_0^* \frac{\delta \G}{\delta \phi_a^*}
                      + \frac{1}{2} \epsilon_{abc} \phi_c^* \frac{\delta \G}{\delta \phi_b^*}
                      - \frac{1}{2} \phi_0^* \frac{\delta \G}{\delta \phi_a^*} = - \Lambda^{D-4} \frac{1}{2} K_0 \phi_a
\label{lfe.1}
\end{eqnarray}
\item the Landau gauge equation
\begin{eqnarray}
\frac{\delta \G}{\delta B_a} =  \Lambda^{D-4}(D^\mu[V](A-V)_\mu)_a \, 
\label{b.eq1}
\end{eqnarray}
and the ghost equation (which holds as a consequence of the ST identity and the Landau gauge equation)
\begin{eqnarray}
\frac{\delta \G}{\delta \bar c_a} = - D^\mu_{ab}[V] \frac{\delta \G}{\delta A^*_{b\mu}}
+ \Lambda^{D-4} (D^\mu[A]\Theta_\mu)_a \, .
\label{gh.1}
\end{eqnarray}
\end{itemize}

Physical unitarity follows from the
ST identity (\ref{st.1}) \cite{Ferrari:2004pd}.

Since
\begin{eqnarray}
\frac{\delta \G^{(0)}}{\delta K_0} = \Lambda^{D-4}
\phi_0 = \Lambda^{D-4} \Big ( v - \frac{1}{2} \frac{\phi_a^2}{v} + \dots
\Big )
\label{invert.1}
\end{eqnarray}
is invertible, the LFE (\ref{lfe.1}) fixes the dependence 
of $\G$ on the $\phi$'s once the ancestor amplitudes
are known, order by order in the loop expansion.
Explicit integration  techniques for the LFE, to all
orders in the loop expansion, have been studied
in \cite{Bettinelli:2007kc}.

\section{Batalin-Vilkovisky bracket}\label{sec:3}

In order to elucidate the meaning of the hierarchy
in terms of canonical transformations we need to make
use of the Batalin-Vilkovisky (BV) formalism
\cite{Batalin:1981jr,Gomis:1994he}
for the model at hand. This requires 
to introduce the
antifields $V_{a\mu}^*, \Theta_{a\mu}^*$ 
paired with $V_a^\mu, \Theta_a^\mu$ and
the antifields $\bar c^*_a, B^*_a$ paired with 
$\bar c_a,B_a$. Moreover one also needs
the antifield $K_0^*$ paired with $K_0$ and
the field $\phi_0^{**}$ 
paired with the antifield $\phi_0^*$. 
$\phi_0^{**}$ is needed because in the nonlinear
theory $\phi_0$ is not an elementary field.

The BV bracket is defined according to the 
conventions of \cite{Gomis:1994he} 
\begin{eqnarray}
\!\!\!\!\!
(X,Y) = \int d^Dx \, \sum_I \Big [
(-1)^{\epsilon_{\Phi_I} (\epsilon_X + 1)}
\frac{\delta X}{\delta \Phi_I}
\frac{\delta Y}{\delta \Phi_I^*} 
- (-1)^{\epsilon_{\Phi_I^*} (\epsilon_X + 1)} 
\frac{\delta X}{\delta \Phi^*_I} 
\frac{\delta Y}{\delta \Phi_I} \Big ] \, .
\label{bv.1.3}
\end{eqnarray}
$\Phi_I, \Phi_I^*$ is a collective notation for the
fields $\{ A_{a\mu}, \phi_a, c_a, V_{a\mu}, \Theta_{a\mu},
\bar c_a, B_a, K_0, \phi_0^{**} \}$ and antifields 
$\{ A^*_{a\mu}, \phi_a^*, c_a^*,
V^*_{a\mu}, \Theta^*_{a\mu}, \bar c^*_a, B^*_a, K^*_0, 
\phi_0^{*} \}$ respectively.
$\epsilon_x$ denotes the statistics of $x$ ($0$ for bosons,
$1$ for fermions). We always use left derivatives.

The couplings of  $\bar c^*_a, V^*_{a\mu}$
and $\phi_0^{**}$ are fixed by the BRST transform of
their partner in the second line of eq.(\ref{st.1}).
This leads us to consider the following tree-level
vertex functional
\begin{eqnarray}
\widehat \G^{(0)} = \left . \G^{(0)} \right |_{\tiny \bar c = B = V_\mu= \Theta_\mu = K_0=\phi_0^*= 0} 
+ \Lambda^{D-4} \int d^Dx \Big (
- \bar c_a^* B_a + V_{a\mu}^* \Theta_a^\mu - K_0 \phi_0^{**}
 \Big ) \, .
\label{bv.1.1}
\end{eqnarray}
The ghost number is assigned as follows. $A_{a\mu},
\phi_a, V_{a\mu}, B_a, K_0, \phi_0^{**}$ have
ghost number zero, $c_a, \Theta_{a\mu}$
have ghost number one, $\bar c_a$ and all
the other antifields with the exception of $c^*_a,
\Theta^*_{a\mu}$ have ghost number $-1$,
while $c^*_a$ and $\Theta^*_{a\mu}$ have ghost
number $-2$. $\widehat \G^{(0)}$ has
ghost number zero.

A canonical transformation (i.e. a transformation
preserving the BV bracket in eq.(\ref{bv.1.3})) 
connects $\widehat \G^{(0)}$ to the original
tree-level vertex functional in eq.(\ref{os.eq.1}).

In order to prove this result
it is convenient to use finite
canonical transformations of the second type \cite{Troost:1989cu}.
They are obtained from a fermionic generating functional
$F(\Phi,\Phi^{'*})$ depending on the old fields $\Phi$ and
the transformed antifields $\Phi^{'*}$ according to
%
%
\begin{eqnarray}
\Phi'_I = \frac{\delta F(\Phi,\Phi^{'*})}{\delta \Phi^{'*}_I} \, , ~~~~~~~~
\Phi^*_I = \frac{\delta F(\Phi,\Phi^{'*})}{\delta \Phi_I} \, .
\label{can.1}
\end{eqnarray}
%
%
%
Then the generating functional of the canonical transformation
by which  $\G^{(0)}$ is recovered from $\widehat \G^{(0)}$ 
(upon setting in the end
 $V^*_{a\mu} =  \phi_0^{**} = \bar c_a^*=0$)
is given by 
\begin{eqnarray}
{\cal F} = \int d^Dx \, \Big ( \phi_0^{*'} (  \phi_0^{**} + \phi_0) 
+ \bar c_a ( \bar c^{*'}_a  -D[V]_\mu (A^\mu - V^\mu)_a ) \Big ) \, ,
\label{bv.2.2}
\end{eqnarray}
where the prime denotes the new variables. 
We do not explicitly write in eq.(\ref{bv.2.2}) the
obvious terms yielding the identity transformation
on the relevant fields and antifields.

The second term in eq.(\ref{bv.2.2}) is the usual gauge-fixing
generating functional (in the background Landau gauge)
\cite{Gomis:1994he}. The first term
takes into account the necessity of introducing a source
for the nonlinear constraint in order to formulate the LFE.

%



The ST identity for $\widehat \G^{(0)}$ can be finally written as
\begin{eqnarray}
{\cal S}(\widehat \G^{(0)}) = \frac{1}{2\Lambda^{D-4}} 
(\widehat \G^{(0)}, \widehat \G^{(0)}) = 0 \, .
\label{bv.1.2}
\end{eqnarray}
This is the master equation \cite{Batalin:1981jr,Gomis:1994he} of the nonlinear theory. 


\section{Quantum and Classical Linearized BV Brackets}\label{sec:4}

We denote by $\widehat \G$ the effective action 
containing the Feynman rules of the theory (tree-level plus
counterterms) 
\begin{eqnarray}
\widehat \G = \sum_{j=0}^{\infty} \widehat \G^{(j)} \, .
\label{ccan.1}
\end{eqnarray}
Since the theory is non-anomalous and we assume to work
in a symmetric regularization scheme, the effective action $\widehat \G$ obeys the master equation 
\begin{eqnarray}
(\widehat \G, \widehat \G) = 0 \, .
\label{ccan.2}
\end{eqnarray}
The operator
%
${\cal S}_{\wG} = (\widehat \G , \cdot)$
%
is nilpotent
\begin{eqnarray}
{\cal S}_{\wG}^2 = 0 \, .
\label{nilp}
\end{eqnarray}
This follows from the master equation
(\ref{ccan.2}) and the (graded) Jacobi identity
for the BV bracket
\cite{Gomis:1994he}
\begin{eqnarray}
((X,Y),Z) + (-1)^{(\epsilon_X+1)(\epsilon_Y + \epsilon_Z)}
((Y,Z),X) + (-1)^{(\epsilon_Z+1)(\epsilon_X + \epsilon_Y)}
((Z,X),Y)=0 \, .
\label{jacobi}
\end{eqnarray}
${\cal S}_{\wG}$ can be filtered
w.r.t. the number of loops
\begin{eqnarray}
{\cal S}_{\wG} = \sum_{j=0}^\infty  {\cal S}_j \, , 
~~~~ {\cal S}_j =  (\wG^{(j)},\cdot)  \, .
\label{n.can.2}
\end{eqnarray}
The quantum BV master equation (\ref{ccan.2})  can be cast
as follows
\begin{eqnarray}
{\cal S}_{\wG} \wG = 0 \, .
\label{n.can.3}
\end{eqnarray}
Notice that 
the lowest order operator ${\cal S}_0$ 
in eq.(\ref{n.can.2})
is also
nilpotent. This can be seen either  by using
eq.(\ref{bv.1.2}) and the Jacobi identity (\ref{jacobi})
or by taking the lowest order in the expansion
of eq.(\ref{nilp}) according to the loop number.

%
%

This allows to define a mapping
${\cal R}$  between the cohomology
classes  
$[X]~\in~H_0({\cal S}_{\wG})$ and $[X^{(0)}] \in 
H_0({\cal S}_0)$ 
at zero ghost number, where $X = \sum_{j=0}^\infty X^{(j)}$
is a local function with ghost number zero graded according to the loop number.
$X^{(j)}$ denotes the coefficient of order $j$ in such an expansion.

The cohomology classes $[X]$ of a nilpotent differential operator $\delta$ are defined by the equivalence
relation
\begin{eqnarray}
X \sim Y \Leftrightarrow \delta X = 0 \, , ~ \delta Y = 0 \, , ~
X = Y + \delta Z \, 
\label{equiv.class}
\end{eqnarray}
for some functional $Z$.
$\delta$ is assumed to increase the ghost number
by one, as is the case for ${\cal S}_{\wG}$
and ${\cal S}_0$.
If $X$ and $Y$ have ghost number zero (and thus
$Z$ has ghost number $-1$), we speak of
the cohomology $H_0(\delta)$ in zero ghost number.
The equivalence class of $X$ is denoted by $[X]$ 
whenever it is clear to which operator the cohomology class
must be referred.

We set
\begin{eqnarray}
{\cal R}[X] = [X^{(0)}] \, .
\label{n.can.5}
\end{eqnarray}
The mapping ${\cal R}$ is well-defined in cohomology, i.e.
${\cal R}[0_{H_0({\cal S}_{\wG})}] = [0_{H_0({\cal S}_0)}]$,
where $[0_{H_0({\cal S}_{\wG})}]$,
$[0_{H_0({\cal S}_0)}] $ is the null cohomology class of $H_0({\cal S}_{\wG})$ resp. $H_0({\cal S}_0)$.
In fact by expanding $X = {\cal S}_{\wG} Y$  according
to the loop number one finds
\begin{eqnarray}
X = X^{(0)} + X^{(1)} + \dots = 
             ({\cal S}_0 + {\cal S}_1 + \dots )
             (Y^{(0)} + Y^{(1)} + \dots) \, .
\label{n.can.6}
\end{eqnarray}
Therefore at lowest order one gets 
$X^{(0)} = {\cal S}_0 Y^{(0)}$ and thus
\begin{eqnarray}
{\cal R}[X] = [{\cal S}_0 Y^{(0)}] = [0_{H_0({\cal S}_0)}] \, .
\label{n.can.7}
\end{eqnarray}
The mapping ${\cal R}$ is an isomorphism. This can be proven
by using standard methods in homological perturbation theory
\cite{homol,Quadri:2002nh}. A short proof of this result is sketched 
in Appendix~\ref{app.proof}.

\section{Classifying Physical Observables}
\label{sec:5}

Since $H_0({\cal S}_{\wG})$ is isomorphic to $H_0({\cal S}_0)$,
the computation of $H_0({\cal S}_0)$ is sufficient in
order to classify the local physical operators of the theory.
In order to carry out this task, we first notice that 
the perturbation theory based on the tree-level vertex
functional in eq.(\ref{bv.1.1}) coincides with the one generated by $\G^{(0)}$
in eq.(\ref{os.eq.1}),
once the canonical transformation induced by
the functional ${\cal F}$ in eq.(\ref{bv.2.2}) 
is performed.

In fact the dependence on
$\bar c_a^*, V^*_{a\mu}, \phi_0^{**}$ is confined
at tree-level due to the validity of the 
following identities for the vertex functional $\G$
\begin{eqnarray}
\frac{\delta \G}{\delta \bar c^*_a} = - \Lambda^{D-4} B_a \, , 
~~~~ \frac{\delta \G}{\delta V^*_{a\mu}} = 
\Lambda^{D-4}  \Theta^\mu_a  \, , ~~~~
\frac{\delta \G}{\delta \phi_0^{**}} = - \Lambda^{D-4}  K_0 \, .
\label{loop.1}
\end{eqnarray}


Since $\G^{(n)}$, $n\geq 1$ does not depend 
on $\bar c^*_a, V^*_{a\mu}, \phi_0^{**}$ (as a consequence
of eq.(\ref{loop.1})) 
and on $\Theta^*_{a\mu}, B^*_a, K_0^*$
(since they do not enter into $\widehat \G^{(0)}$) 
we can limit ourselves to the local functional space spanned
by $\{ A_{a\mu}, \phi_a, c_a, V_{a\mu}, \Theta_{a\mu},
\bar c_a, B_a, K_0 \}$ and  
$\{ A^*_{a\mu}, \phi_a^*, c_a^*, \phi_0^{*} \}$. 

It is convenient to  introduce 
a matrix notation and set
\begin{eqnarray}
A_\mu = A_{a\mu} \frac{\tau_a}{2} \, ,
\label{Ncan.2}
\end{eqnarray}
where $\tau_a$ are Pauli matrices.
The Goldstone fields $\phi_a$ and the nonlinear constraint
$\phi_0$ are gathered into the SU(2) matrix
\begin{eqnarray}
&& \Omega=\frac{1}{v} ( \phi_0 + i \tau_a \phi_a) \, , ~~~~
\Omega^\dagger \Omega = 1 \, ,~~~~ {\rm det} ~ \Omega = 1 \, , \nonumber \\
&& \phi_0^2 + \phi_a^2 = v^2 \, .
\label{Ncan.3}
\end{eqnarray}
The SU(2) flat connection is defined in terms of 
$\Omega$ by
\begin{eqnarray}
&& F_\mu = F_{a\mu} \frac{\tau_a}{2} = i \Omega \partial_\mu \Omega^\dagger \, . 
\label{Ncan.4}
\end{eqnarray}
$F_{a\mu}$ reads in components
\begin{eqnarray}
F_{a\mu} = \frac{2}{v^2} ( \phi_0 \partial_\mu \phi_a -
\partial_\mu \phi_0 \phi_a + \epsilon_{abc} \partial_\mu \phi_b 
\phi_c ) \, .
\label{Neq.flatconn.comp}
\end{eqnarray}

A finite SU(2)$_L$ gauge transformation acts as follows:
\begin{eqnarray}
&& \Omega' = U_L \Omega \, , \nonumber \\
&& A'_\mu = U_L A_\mu U_L^\dagger + i U_L \partial_\mu U_L^\dagger \, , \nonumber \\
&& F'_\mu = U_L F_\mu U_L^\dagger + i U_L \partial_\mu U_L^\dagger \, .
\label{Ncan.5}
\end{eqnarray}

The computation of the cohomology
$H_0({\cal S}_0)$ is simplified if one moves
to gauge-invariant (bleached) variables,
which automatically satisfy the classical
linearized LFE \cite{Bettinelli:2007tq}.

From eq.(\ref{Ncan.5})
one sees that the following combination is invariant
under a local SU(2)$_L$ transformation
\begin{eqnarray}
a_\mu = \Omega^\dagger (A_\mu - F_\mu) \Omega = 
\Omega^\dagger A_\mu \Omega - i \Omega^\dagger \partial_\mu \Omega \, .
\label{Ncan.6}
\end{eqnarray}
We call $a_\mu$ the bleached counterpart of the original
gauge connection $A_\mu$. 

The bleached counterpart of 
the background connection $V_\mu$ is 
\begin{eqnarray}
v_\mu = \Omega^\dagger (V_\mu - F_\mu) \Omega = 
\Omega^\dagger V_\mu \Omega - i \Omega^\dagger \partial_\mu \Omega \, .
\label{Ncan.7}
\end{eqnarray}
The bleached counterparts of the ghost field
$c = c_a \frac{\tau_a}{2}$, the ghost background source
$\Theta_\mu = \Theta_{a\mu} \frac{\tau_a}{2}$,
the ghost antifield $c^* = c^*_a \frac{\tau_a}{2}$
are defined by
\begin{eqnarray}
&& \tilde c = \Omega^\dagger c \Omega \, , ~~~~ 
      \tilde \Theta_\mu = \Omega^\dagger \Theta_\mu \Omega \, ,  ~~~~
      \tilde c^* = \Omega^\dagger c^* \Omega \, .
\label{Ncan.8}
\end{eqnarray}
By exploiting the ghost equation (\ref{gh.1}) or alternatively by performing
the canonical transformation in eq.(\ref{bv.2.2}) one sees that 
the vertex functional $\G$ only depends on the
combination
\begin{eqnarray}
 \widehat A^*_{a\mu} = A^*_{a\mu} + (D_\mu[V]\bar c)_a \, .
\label{Ncan.9}
\end{eqnarray}
The bleached counterpart of $\widehat A^*_\mu =  \widehat A^*_{a\mu}
\frac{\tau_a}{2}$ is
\begin{eqnarray}
\widetilde{\widehat  A^*_\mu} = \Omega^\dagger \widehat  A^*_\mu \Omega \, .
\label{Ncan.10}
\end{eqnarray}
Unlike in \cite{Bettinelli:2007tq} we do not work with the bleached variables of $\phi^*_a$.
In fact the canonical transformation in eq.(\ref{bv.2.2}) generates
the combination
\begin{eqnarray}
\widehat \phi_a^* = \phi_a^* - \frac{\phi_a}{\phi_0} \phi_0^* 
\label{Ncan.11}
\end{eqnarray}
Then by explicit computation one finds that its ${\cal S}_0$-variation is
\begin{eqnarray}
{\cal S}_0 \widehat \phi_a^*  = \left . \frac{\delta \G^{(0)}}{\delta \phi_a}
\right |_{K_0=0} 
\label{Ncan.12}
\end{eqnarray}
where the R.H.S. is expressed as a function of $\widehat \phi_a^*$. I.e.
the canonical transformation in eq.(\ref{bv.2.2}) allows to recover
precisely the tangent space of the group SU(2).
This geometrical property is of course expected
and becomes transparent in the approach based
on the BV formalism. Moreover we notice that $K_0$ and $\phi_0^*$ form a ${\cal S}_0$-doublet  \cite{Quadri:2002nh}
and consequently they do not enter into the non-trivial
cohomology classes of $H({\cal S}_0)$.

By direct computation one then obtains  the ${\cal S}_0$-transforms of the other variables 
\cite{Bettinelli:2007tq}:
\begin{eqnarray}
&& {\cal S}_0 a_\mu = 0 \, , ~~~~~  {\cal S}_0 \tilde c = -\frac{i}{2}
\{ \tilde c, \tilde c \} \, ,
\nonumber \\
&& {\cal S}_0 v_\mu = \tilde \Theta_\mu - D_\mu[v] \tilde c \, , ~~~~~  
      {\cal S}_0 \tilde \Theta_\mu = - i \{ \tilde c, \tilde \Theta_\mu \} \, , \nonumber \\
&& {\cal S}_0 \widetilde{\widehat A^*_\mu} = \Lambda^{D-4}
\Big [ \frac{1}{g^2} D^\rho G_{\rho\mu}[a] + M^2 a_\mu \Big ] \, , \nonumber \\
&& {\cal S}_0 \tilde c^* = (D^\mu[a] \widetilde{\widehat A^*}_\mu) - 
\frac{i}{4} {\Omega^*}^\dagger \Omega + \frac{i}{8} T_r [{\Omega^*}^\dagger \Omega ] {\bf 1} \, .
\label{table.1}
\end{eqnarray}
$\Omega^*$ is a $2\times2$ matrix defined by
\begin{eqnarray}
\Omega^* = \phi_0^* + i \phi_a^* \tau_a \, .
\label{omegast}
\end{eqnarray}
The combinations
\begin{eqnarray}
 {\tilde \Theta}'_\mu = \tilde \Theta_\mu - D_\mu[v] \tilde c \, ,
~~
-\frac{i}{4} {\Omega^*}^{\dagger '} \equiv
-\frac{1}{4} \phi_a^{*'} \tau_a = 
(D^\mu[a] \widetilde{\widehat A^*}_\mu) - 
\frac{i}{4} {\Omega^*}^\dagger \Omega + \frac{i}{8} T_r [{\Omega^*}^\dagger \Omega ] {\bf 1}
\label{s0doub}
\end{eqnarray}
form ${\cal S}_0$-doublets with $v_\mu$
and $\tilde c^*$ respectively.
Moreover the change of variables
$\tilde \Theta^\mu_a \rightarrow \tilde \Theta^{'\mu}_a$,
$\phi_a^* \rightarrow \phi_a^{*'}$ is invertible.
Thus the pairs $(v_\mu, \tilde \Theta'_\mu)$,
$(\tilde c^*, -\frac{i}{4} {\Omega^*}^{\dagger '})$ 
cannot contribute to the non-trivial cohomology
classes of $H_0({\cal S}_0)$.
Hence the latter only depend on
$a_\mu, \tilde c, \widetilde{\widehat{A_\mu^*}}$.

The cohomology $H_0({\cal S}_0)$ in this space
has been computed in \cite{Bettinelli:2007tq,Henneaux:1998hq} 
and is given by all possible local polynomials 
built out from $a_\mu$ and its ordinary derivatives modulo the ideal generated by the 
transformation of $\widetilde{\widehat A^*_\mu}$ (which yields the classical equation of motion for the gauge field $a_\mu$) plus cohomologically trivial ${\cal S}_0$-exact terms.
In particular the whole dependence on all variables but $a_\mu$ is confined to the
${\cal S}_0$-exact sector.

Since the theory is also invariant under global
$SU(2)_R$ symmetry, 
only global $SU(2)_R$-invariant operators
need to be considered.

\medskip
The construction of bleached variables is not limited to the fields
and antifields of pure Yang-Mills theory.
As an example, for a fermion matter doublet $L$ transforming in the fundamental representation of SU(2)
\begin{eqnarray}
L' = U_L L \, 
\label{eq.7}
\end{eqnarray}
its bleached counterpart is 
\begin{eqnarray}
\tilde L = \Omega^\dagger L \, .
\label{eq.8}
\end{eqnarray}
The construction can be generalized to fields in arbitrary representations
of the gauge group along the lines of \cite{Henneaux:1998hq}.

\section{Canonical Transformations For The Bleached Variables}
\label{sec:6}

It remains to be shown that the bleached variables 
discussed in the previous Section can indeed be obtained
via a canonical transformation. For that purpose the easiest
way is to provide  the relevant generating functional,
which looks as follows
\begin{eqnarray}
{\cal F}_1 & = & \int d^Dx \, \Bigg (
 2 ~ T_r  [  \widetilde{\widehat A^{*}_{\mu}}  a^\mu(\vec{\phi},A_\nu) ] 
+ 2 ~ T_r [ \widetilde c^{*} \tilde c(\vec{\phi},c) ] \nonumber \\
& & ~~~~~~~~~ + 2 ~ T_r [ \widetilde V^{*}_{\mu} v^\mu(\vec{\phi},V_\nu)] +
2 ~ T_r [ \widetilde \Theta^{*}_\mu \tilde \Theta^{\mu}(\vec{\phi}, \Theta) ]  
 \Bigg ) \, .
\label{can.9}
\end{eqnarray}
The associated canonical transformation automatically induces the 
bleaching transformation on the antifields. 
Moreover the redefinition of $\phi_a^*$ 
is  the one required in order to generate the transformation
properties of the bleached variables, as expected. 

As an example, let us work out in detail
the term proportional to $\Astarbl$ of such a redefinition.
One finds
\begin{eqnarray}
\phi_a^{*}(z) & = & \frac{\delta {\cal F}_1}{\delta \phi_a(z)} 
\nonumber \\
& = & \frac{\delta}{\delta \phi_a(z)} 
2 \int d^Dx \, T_r [\Astarbl ( i \Omega^\dagger \partial^\mu \Omega
+ \Omega^\dagger A^\mu \Omega)]
\nonumber \\
& = & 2  \int d^Dx \,
T_r [ \Astarbl (
i \frac{\delta \Omega^\dagger}{\delta \phi_a(z)} \partial^\mu \Omega
+ i \Omega^\dagger \frac{\delta}{\delta \phi_a(z)} \partial^\mu \Omega
\nonumber \\
& & ~~~~~~~~~~~~~~~~~~~ +
\frac{\delta \Omega^\dagger}{\delta \phi_a(z)} A^\mu \Omega +
\Omega^\dagger A^\mu \frac{\delta \Omega}{\delta \phi_a(z)} ) ] \nonumber \\
& = &  2 \int d^Dx \,
T_r [ \Astarbl (
i \frac{\delta \Omega^\dagger}{\delta \phi_a(z)} \partial^\mu \Omega
+ i \Omega^\dagger \frac{\delta}{\delta \phi_a(z)} \partial^\mu \Omega
\nonumber \\
& & ~~~~~~~~~~~~~~~~~~~ +
\frac{\delta \Omega^\dagger}{\delta \phi_a(z)} 
( \Omega a_\mu \Omega^\dagger + i \Omega \partial_\mu \Omega^\dagger )
\Omega \nonumber \\
& & ~~~~~~~~~~~~~~~~~~~ +
\Omega^\dagger
( \Omega a_\mu \Omega^\dagger + i \Omega \partial_\mu \Omega^\dagger )
\frac{\delta \Omega}{\delta \phi_a(z)} ) ] \, .
\label{new.can.6}
\end{eqnarray}
By using the nonlinear constraint one gets
\begin{eqnarray}
0 = \frac{\delta}{\delta\phi_a} (\Omega^\dagger \Omega) = 
\frac{\delta \Omega^\dagger}{\delta \phi_a} \Omega +
\Omega^\dagger \frac{\delta \Omega}{\delta \phi_a} 
\label{new.can.7}
\end{eqnarray}
By using eq.(\ref{new.can.7}) into eq.(\ref{new.can.6}) we 
end up with
\begin{eqnarray}
\phi_a^{*}(z) & = & 2 \int d^Dx \, 
T_r [ \Astarbl ( i \Omega^\dagger \frac{\delta}{\delta \phi_a(z)} 
\partial^\mu \Omega + i \partial^\mu \Omega^\dagger 
\frac{\delta \Omega}{\delta \phi_a} +
[ a_\mu, \Omega^\dagger \frac{\delta \Omega}{\delta \phi_a(z)} ] ) ] 
\nonumber \\
& = & 2i \int d^Dx \, 
T_r [ \Astarbl D^\mu[a] ( \Omega^\dagger \frac{\delta \Omega}{\delta \phi_a(z)})
] \, . 
\label{new.can.8}
\end{eqnarray}
This has to be inserted back into the piece
\begin{eqnarray}
\int d^Dx \, \phi_a^* \s\phi_a 
\label{int.1}
\end{eqnarray}
of the tree-level vertex functional.
By plugging eq.(\ref{new.can.8}) into eq.(\ref{int.1}) one 
finally obtains
\begin{eqnarray}
&& 2i \int d^Dx \, T_r [ \Astarbl D^\mu[a] \int d^Dz \, 
(\Omega^\dagger \s\phi_a(z) \frac{\delta \Omega}{\delta \phi_a(z)})] \nonumber \\
&& ~~~~~ = 2i
 \int d^Dx \, T_r [ \Astarbl D^\mu[a] \int d^Dz \, 
(\Omega^\dagger \s \Omega)] \nonumber \\
&& ~~~~~ = -2 \int d^Dx \, T_r [ \Astarbl D^\mu[a] 
\Omega^\dagger \Omega \tilde c] \nonumber \\
&& ~~~~~ = - 2 \int d^Dx \, T_r [ \Astarbl D^\mu[a] 
 \tilde c] 
\label{int.2}
\end{eqnarray}
which exactly cancels the $\widetilde{\widehat A^*}$-dependent term in $\widehat \G^{(0)}$. This cancellation corresponds to the fact
that the bleached variable $a_\mu$ is 
${\cal S}_0$-invariant and therefore there should be no dependence on its antifield
in $\widehat \G^{(0)}$ (after the canonical
transformation generated by ${\cal F}_1$ is implemented).

\section{Comparison With The Linear Theory}
\label{sec:7}

As it has been shown in Sect.~\ref{sec:5}, the physical observables of the theory are classified by
$H_0({\cal S}_0)$, which in turn is given by
all possible global
$SU(2)_R$-invariant local polynomials in $a_\mu$ 
and ordinary derivatives thereof which do not vanish
on the classical equation of motion of $a_\mu$.

%
The requirement of the validity of the WPC  selects 
the Yang-Mills field strength squared and the 
St\"uckelberg mass term as the only possible physical
operators admissible in the tree-level vertex functional
\cite{Bettinelli:2007tq}.

A comparison with the linearly realized SU(2) 
Yang-Mills theory can be useful. In the linearly
realized framework the trace component $h$
of the $2 \times 2$ matrix
\begin{eqnarray}
H = h + i \phi_a \tau_a \, , ~~~ h = v+ \sigma
\label{lin.1}
\end{eqnarray}
is an independent degree of freedom.  The latter is parameterized
by the Higgs field $\sigma$.
 $h$ acquires
the vacuum expectation value $v$ via 
spontaneous symmetry breaking and correspondingly $\langle \sigma \rangle=0$.

The construction of gauge-invariant variables out of $A_{a\mu}$
and $\sigma$ is easily performed via the field redefinitions
\begin{eqnarray}
&& A_{a\mu} \rightarrow \tilde h_{a\mu} = T_r \Big \{
\frac{i}{H^\dagger H} H^\dagger D_\mu[A] H \tau_a \Big \} \, , 
\label{lin.2.1} \\
&& \sigma \rightarrow \tilde \sigma = \sqrt{H^\dagger H} - v \, ,
\label{lin.2.2}
\end{eqnarray}
where $D_\mu[A]$ is the covariant derivative
\begin{eqnarray}
D_\mu[A] = \partial_\mu - i A_{a\mu} \frac{\tau_a}{2} \, .
\label{lin.3}
\end{eqnarray}
Since $H$ tranforms in the fundamental representation of SU(2)
\begin{eqnarray}
H' = U_L H \, , 
\label{lin.4}
\end{eqnarray}
the R.H.S. of eqs.(\ref{lin.2.1}),(\ref{lin.2.2}) are automatically
gauge invariant.
In fact eqs.(\ref{lin.2.1}), (\ref{lin.2.2}) can be understood
as the result of an operatorial finite gauge transformation,
generated by the matrix $H^\dagger / \sqrt{H^\dagger H} \in \mbox{SU(2)}$, acting on $A_\mu$ and $H$ respectively.
 At $\phi_a=0$ $\tilde h_{a\mu}$ and
$\tilde \sigma$ reduce to $A_{a\mu}$ and $\sigma$.
In the linearly realized theory $\sigma$ is an ancestor field.

\begin{figure}
\begin{center}
\includegraphics[width=3truecm]{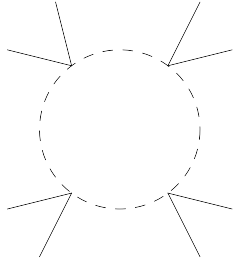}
\end{center}
\caption{One-loop divergent graph with arbitrary
number of external Higgs legs. Dashed lines denote
Goldstone propagators (in the 't Hooft gauge).}
\label{fig.1}
\end{figure}

Any functional built out of $\tilde h_{a\mu}, \tilde \sigma$
and ordinary derivatives thereof is gauge-invariant. 
As a consequence, the following mass bilinears
\begin{eqnarray}
m_{ab} \tilde h_{a\mu} \tilde h_b^\mu \, , ~~~~ m_{ab}= m_{ba}
\label{lin.5}
\end{eqnarray}
are admissible on symmetry grounds.
However, upon expansion of $\tilde h_{a\mu}$ in components
\begin{eqnarray}
\tilde h_{a\mu} & = & A_{a\mu} - \frac{2}{v} \Big ( \partial_\mu \phi_a
+ \epsilon_{abc} A_{b\mu} \phi_c \Big ) \nonumber \\
&& 
+ \frac{2}{v^2} \Big ( \sigma \partial_\mu \phi_a +
\phi_a \partial_\mu \sigma - A_{a\mu} \vec{\phi}^2 
- \epsilon_{abc} \phi_b \partial_\mu \phi_c \Big ) + O(1/v^3)
\label{lin.6}
\end{eqnarray}
one sees that eq.(\ref{lin.5}) contains vertices involving two $\sigma$'s,
two $\phi$'s and two derivatives. Thus at one loop level
diagrams like those in Fig.~\ref{fig.1} arise.
They are logarithmically divergent irrespective of the number of
external $\sigma$-legs.
A similar argument shows that the kinetic term
$\partial_\mu \tilde \sigma \partial^\mu \tilde \sigma$
contains a vertex $\sim \frac{1}{v^2} \sigma \partial_\mu \sigma 
\phi_a \partial^\mu \phi_a$, which gives rise to the same diagrams
as in Fig.~\ref{fig.1}.
This implies that the WPC is maximally violated in the linear theory,
unless one chooses the combination
\begin{eqnarray}
T_r \Big \{  H^\dagger H ~ \frac{(-i)}{H^\dagger H} (D_\mu[A] H)^\dagger H 
\frac{i}{H^\dagger H} H^\dagger D^\mu[A] H \Big \} = 
T_r \{ (D_\mu[A] H)^\dagger D^\mu[A] H \}\, ,
\label{lin.7}
\end{eqnarray}
i.e. the WPC in the linear theory is strong enough to
select a single gauge-invariant combination  
which boils down to the usual covariant kinetic term
(\ref{lin.7})
for the Higgs doublet $H$.

\section{SU(2) $\otimes$ U(1)} \label{sec:8}

For the EW group SU(2) $\otimes$ U(1) the SU(2) custodial symmetry
\cite{Sikivie:1980hm} is violated in the fermionic sector. In the nonlinearly
realized theory this fact entails that two independent gauge bosons mass invariants
can be introduced \cite{Bettinelli:2008ey}
in a way compatible with the WPC. They can be parameterized
as
\begin{eqnarray}
M^2 Tr \Big \{ (g A_\mu - \frac{g'}{2} \Omega \tau_3 B_\mu \Omega^\dagger - F_\mu)^2 \} +
M^2 \frac{\kappa}{2} \Big ( Tr \{ ( (g A_\mu - \frac{g'}{2} \Omega \tau_3 B_\mu \Omega^\dagger - F_\mu) \tau_3 \} \Big )^2 \, .
\label{2mi}
\end{eqnarray}
$g,g'$ are the SU(2) and U(1)$_Y$ coupling constants
repectively, $B_\mu$ the U(1)$_Y$ connection.
Notice that  we have restored the coupling constants in front of the
gauge fields in order to match with the conventions
of \cite{Bettinelli:2008ey}.
The action of U(1)$_Y$ on $\Omega$ is on the right, i.e.
\begin{eqnarray}
\Omega ' = \Omega V^\dagger \, , ~~~~~~ V=\exp \Big ( i \alpha
\frac{\tau_3}{2} \Big ) \, .
\label{u1.y}
\end{eqnarray}
One can introduce the bleached (SU(2) invariant) combination
\cite{Bettinelli:2008ey}
\begin{eqnarray}
w_\mu = w_{a\mu} \frac{\tau_a}{2} = 
g \Omega^\dagger A_\mu \Omega - g' B_\mu \frac{\tau_3}{2} +
i \Omega^\dagger \partial_\mu \Omega 
\label{ew.bleach}
\end{eqnarray}
Under U(1)$_Y$ one gets
\begin{eqnarray}
w'_\mu = V w_\mu V^\dagger \, .
\label{w.u1}
\end{eqnarray}
Since $w_\mu$ is SU(2)-invariant, by the Gell-Mann-Nishijima
formula one sees that on these variables the action of U(1)$_Y$ coincides
with the action of U(1)$_{em}$. The two independent
bilinears in eq.(\ref{2mi}) correspond to independent
mass terms for the two electrically neutral combinations
\begin{eqnarray}
M^2 \Big ( w^+ w^- + \frac{1}{2} w_3^2 \Big )\, , ~~~~~~ 
\frac{M^2 \kappa}{2} w_3^2 \, .
\label{bleach.inv.mass}
\end{eqnarray}

On the other hand, for the linearly realized EW model the WPC condition
by itself is sufficient in order to impose the validity of the
tree-level Weinberg relation. I.e. relaxing power-counting renormalizability
(in favour of the weaker WPC condition) does not allow to introduce
a second independent mass parameter for the non-Abelian gauge bosons.

The argument closely parallels the one given in Sect.~\ref{sec:6}:
interaction vertices with two $\sigma$'s, two $\phi$'s and two derivatives
only disappear if the combination (\ref{lin.7}) is chosen, where now
the covariant derivative must be replaced with its SU(2) $\otimes$ U(1)
counterpart, i.e.
\begin{eqnarray}
D_\mu H = \partial_\mu H - i g A_{a\mu} \frac{\tau_a}{2} H - i g' H B_\mu
\frac{\tau_3}{2} \, .
\label{cov.1}
\end{eqnarray}
This feature might have some interesting phenomenological
consequences. From the modern point of view which considers
the SM as a very accurate effective approximation to a
more fundamental theory, it makes sense to use the WPC
as a guiding tool for controlling the loop perturbative expansion.

The low-energy
limit of the more fundamental theory endowed with the exact custodial
symmetry can lead to a model where two
independent mass parameters in the vector meson sector
are allowed only if at low energies
the EW symmetry is non-linearly realized.
On the other hand, a stronger remnant (imposing the exact relation
$\kappa=0$) would be in place if the low-energy realization of the EW symmetry
were linear.

If a global fit (including radiative corrections) to EW precision data 
\cite{:2005ema} favours a solution
where $\kappa \neq 0$, this might be an indirect evidence
that at LEP energies the EW symmetry is in fact nonlinearly
realized.

\section{Conclusions}\label{sec:9}

We have classified the physical observables in the
nonlinearly realized massive SU(2) Yang-Mills theory
within the mathematically consistent framework
governed by the LFE and the WPC.
This approach allows to take into account
the non-trivial quantum
deformations of the nonlinearly realized
gauge symmetry to all orders in the loop expansion.

It has been shown that the bleached variables,
introduced in 
\cite{Ferrari:2005va,Bettinelli:2007tq,Bettinelli:2008ey}
 as solution
of the linearized LFE, can be obtained
through a canonical transformation w.r.t. the BV bracket
associated with the BRST symmetry of the model.
In this process the tangent space of the group SU(2)
emerges naturally.

The role of the WPC
in the linear vs. the nonlinear realization of the
gauge symmetry has been clarified.
We have found that the tree-level Weinberg relation
in the EW theory holds even though
power-counting renormalizability is dropped in favour of the weaker WPC.

From the modern 
point of view which considers the SM as a very accurate effective approximation to a 
more fundamental theory, the WPC 
can be used as a unified guiding tool for controlling 
the loop perturbative expansion 
both in the linearly and in the nonlinearly realized EW theory.

One can then compare the global fit to the existing
LEP precision data \cite{:2005ema}.
Should the  solution
with a non-zero
mass parameter $\kappa$ be preferred,
this might represent a rather intriguing indication that the EW
symmetry is in fact nonlinearly realized at the LEP scale. 
This would also point towards a scenario
with no SM Higgs,
an option which could be experimentally
investigated at the LHC in the coming years
\cite{DeRoeck:2009id}.

The fit within
the nonlinear EW model  must face some non-trivial issues. 
Explicit  computations \cite{pip} show in fact that 
the radiative corrections to pseudo-observables at the Z pole
in the nonlinearly realized EW theory are not-oblique \cite{oblique}
and get some flavour-dependent non-SM-like corrections
(via their top mass dependence).
The comparison with the experimental data
must be performed in such a way that the SM-dependent
assumptions, controlling the experimental fit of 
Ref.~\cite{:2005ema}, are
properly taken into account in the non-linear setting, in particular 
in connection with the dependence on the
second mass parameter of the hadronic contribution
to the Z-$\gamma$ interference term \cite{:2005ema}.
This deserves further investigation.

\section*{Acknowledgments}

I thank Glenn Barnich, Alberto Cattaneo, Stefan Dittmaier
and Ruggero Ferrari for very stimulating discussions.
The hospitality of 
the Physics Department at the ULB Bruxelles and of
the Department of Mathematics
at the Z\"urich University 
is gratefully acknowledged.

\appendix
\section{$H_0({\cal S}_{\wG}) \sim H_0({\cal S}_0)$}\label{app.proof}

In this Appendix we prove that the mapping ${\cal R}$ 
between $H_0({\cal S}_{\wG})$ and $H_0({\cal S}_0)$ is
an isomorphism.

In order to show that ${\cal R}$ is one-to-one we prove
that $\mbox{ker} ~ {\cal R} = \{ [ 0_{H_0({\cal S}_0)} ] \}$.
If $[{\cal I}] \in \mbox{ker} ~ {\cal R}$, then its lowest order
coefficient ${\cal I}^{(0)}$ is ${\cal S}_0$-exact, i.e.
\begin{eqnarray}
{\cal I}^{(0)} = {\cal S}_0 {\cal G}_0
\label{app.1}
\end{eqnarray}
for some local function ${\cal G}_0$. Then one
can write
\begin{eqnarray}
{\cal I}  =  {\cal I}^{(0)} + {\cal I} - {\cal I}^{(0)} & = & {\cal S}_0 {\cal G}_0 +
{\cal I} - {\cal I}^{(0)} \nonumber \\
& = & {\cal S}_{\wG} {\cal G}_0 + {\cal H}_1 
\label{app.2}
\end{eqnarray}
where
\begin{eqnarray}
{\cal H}_1 = - ({\cal S}_{\wG} - {\cal S}_0) {\cal G}_0 + {\cal I} - {\cal I}^{(0)}
\label{app.3}
\end{eqnarray}
is of order at least one in the loop expansion. 
Let us now suppose that ${\cal I}$ is ${\cal S}_{\wG}$-exact
up to order~$k$:
\begin{eqnarray}
{\cal I} = {\cal S}_{\wG} {\cal G}_{j-1} + {\cal H}_j \, , ~~~~
j=1,2,\dots,k 
\label{app.4}
\end{eqnarray}
${\cal H}_j$ is at least of order $j$ in the loop expansion.
By the nilpotency of ${\cal S}_{\wG}$ one obtains from
eq.(\ref{app.4})
\begin{eqnarray}
{\cal S}_{\wG} {\cal H}_k = 0 \, .
\label{app.5}
\end{eqnarray}
By projecting the above equation at the lowest
non-vanishing order one gets
\begin{eqnarray}
{\cal S}_0 {\cal H}_k^{(k)} = 0 \, .
\label{app.6}
\end{eqnarray}
Since we assume that the theory is non-anomalous, the cohomology of ${\cal S}_0$ is empty in ghost number
one, i.e. there exists a local function
${\cal G}^{(k)}$ such that
\begin{eqnarray}
{\cal H}^{(k)}_k = {\cal S}_0 {\cal G}^{(k)} \, .
\label{app.7}
\end{eqnarray}
Then 
\begin{eqnarray}
{\cal I} & = & {\cal S}_{\wG} {\cal G}_{k-1} + {\cal H}^{(k)}_k + {\cal H}^{(k+1)}_k + \dots \nonumber \\
& = & {\cal S}_{\wG} {\cal G}_{k-1} + {\cal S}_0 {\cal G}^{(k)}_k
+ {\cal H}^{(k+1)}_k + \dots \nonumber \\
& & {\cal S}_{\wG} ( {\cal G}_{k-1} + {\cal G}_k^{(k)} ) -
({\cal S}_{\wG} - {\cal S}_0) {\cal G}^{(k)}_k 
+ {\cal H}_k - {\cal H}_k^{(k)} \nonumber \\
& = & {\cal S}_{\wG} {\cal G}_k + {\cal H}_{k+1} 
\label{app.8}
\end{eqnarray}
where ${\cal G}_k = {\cal G}_{k-1} + {\cal G}^{(k)}_k$ 
and 
\begin{eqnarray}
 {\cal H}_{k+1} = -
({\cal S}_{\wG} - {\cal S}_0) {\cal G}^{(k)}_k 
+ {\cal H}_k - {\cal H}_k^{(k)} 
\label{app.9}
\end{eqnarray}
i.e. ${\cal I}$ is ${\cal S}_{\wG}$-exact up to order $k+1$.

Moreover if
\begin{eqnarray}
{\cal S}_0 {\cal I}^{(0)} = 0 
\label{app.10}
\end{eqnarray}
one can recursively find coefficients
${\cal I}^{(j)}$, $ j \geq 1$ in such a way that
\begin{eqnarray}
{\cal I} = \sum_{j=0}^\infty {\cal I}^{(j)} 
\label{app.11}
\end{eqnarray}
is ${\cal S}_{\wG}$-invariant. 
This can be proven as follows.
Nilpotency of ${\cal S}_{\wG}$ yields at order one
\begin{eqnarray}
{\cal S}_0 {\cal S}_1 + {\cal S}_1 {\cal S}_0 = 0 \, .
\label{app.12}
\end{eqnarray}
By using the above equation one obtains from eq.(\ref{app.10}) 
\begin{eqnarray}
{\cal S}_0 {\cal S}_1 {\cal I}^{(0)} = 0 \, .
\label{app.13}
\end{eqnarray}
Since the cohomology of ${\cal S}_0$ is empty at ghost
number one (no anomalies), there exists a local function
${\cal I}^{(1)}$ such that
\begin{eqnarray}
{\cal S}_1 {\cal I}^{(0)} = - {\cal S}_0 {\cal I}^{(1)} \, .
\label{app.14}
\end{eqnarray}
Therefore 
\begin{eqnarray}
{\cal S}_1 {\cal I}^{(0)} + {\cal S}_1 {\cal I}^{(0)} = 0 \, .
\label{app.15}
\end{eqnarray}
Suppose now that ${\cal S}_{\wG} {\cal I}=0$ holds up to order
$n-1$
\begin{eqnarray}
\sum_{j=0}^m {\cal S}_j {\cal I}^{(m-j)} = 0 \, , ~~~~ m=0,1,\dots,n-1
\label{app.16}
\end{eqnarray}
Then
\begin{eqnarray}
{\cal S}_{\wG} \sum_{k=0}^{n-1} {\cal I}^{(k)} =
\Delta^{(n)} + \dots
\label{app.17}
\end{eqnarray}
where
\begin{eqnarray}
\Delta^{(n)} = \sum_{j=1}^n {\cal S}_j {\cal I}^{(n-j)} \, .
\label{app.18}
\end{eqnarray}
Again by the nilpotency of ${\cal S}_{\wG}$ one gets
\begin{eqnarray}
&& {\cal S}_{\wG} \Delta^{(n)} = 0 \, , ~~~~
({\cal S}_0 + {\cal S}_1 + \dots)(\Delta^{(n)} + \dots) = 0 \, .
\label{app.19}
\end{eqnarray}
The projection of the above equation at lowest order
gives
\begin{eqnarray}
{\cal S}_0 \Delta^{(n)} = 0 
\label{app.20}
\end{eqnarray}
which, again under the assumption that no anomalies
are present, implies
\begin{eqnarray}
\Delta^{(n)} = -{\cal S}_0 {\cal I}^{(n)} 
\label{app.21}
\end{eqnarray}
for a local function ${\cal I}^{(n)}$ of order $n$
in the loop expansion.
Then by eq.(\ref{app.21}) one has
\begin{eqnarray}
{\cal S}_0 {\cal I}^{(n)} + \sum_{j=1}^n
{\cal S}_j {\cal I}^{(n-j)} = 0
\label{app.22}
\end{eqnarray}
i.e. eq.(\ref{app.16}) holds also at order $n$.
This concludes the proof.

%
%
%
%
%

\end{document}